\documentclass[12pt,preprint]{aastex}

\usepackage{graphicx}
\usepackage{times}

\usepackage{longtable}

\usepackage{xspace}

\newcommand{\aastexurl}{}
\aastexurl\url
\let\url\relax

\usepackage{hyperref}
\hypersetup{
            pdftitle={Photometry of {SN~2002ic} and Implications for the Progenitor Mass-loss History},
            pdfauthor={W. Michael Wood-Vasey, UC Berkeley/LBNL and Lifan Wang, LBNL}
            bookmarks=true,
            bookmarksnumbered=true,
            colorlinks=true,
            linkcolor=blue
           }

\begin{document}
\title{Photometry of {SN~2002ic} and Implications for the Progenitor Mass-Loss History}
\shorttitle{Photometry of SN~2002ic and Progenitor Mass-Loss}
\shortauthors{Wood-Vasey et al.}
\author{W.~M. Wood-Vasey\altaffilmark{1} and L. Wang and G. Aldering}
\affil{Physics Division, Lawrence Berkeley National Laboratory, Berkeley, CA 94720}
\altaffiltext{1}{Department of Physics, University of California, Berkeley, CA 94720}
\email{wmwood-vasey@lbl.gov}

\date{2004 May 6}
\received{2004 March 12}

\begin{abstract}
We present new pre-maximum and late-time optical photometry of the
Type~Ia/IIn supernova 2002ic. These observations are combined with the
published V-band magnitudes of~\citet{hamuy03b} and the
VLT spectrophotometry of~\citet{wang04} to construct the most extensive
light curve to date of this unusual supernova.  The observed flux at
late time is significantly higher relative to the flux at maximum than
that of any other observed Type~Ia supernova and continues to fade
very slowly a year after explosion.  Our analysis of the light curve
suggests that a non-Type Ia supernova component becomes prominent
$\sim20$~days after explosion.  Modeling of the non-Type Ia supernova
component as heating from the shock interaction of the supernova
ejecta with pre-existing circumstellar material suggests the presence
of a $\sim1.7\times10^{15}$~cm gap or trough between the progenitor system 
and the surrounding circumstellar material.
This gap could be due to significantly lower mass-loss
$\sim15~(\frac{v_w}{10~\mathrm{km/s}})^{-1}$~years 
prior to explosion or evacuation of the circumstellar material by a
low-density fast wind.  The latter is consistent with observed
properties of proto-planetary nebulae and with models of white-dwarf +
asymptotic giant branch star progenitor systems with the asymptotic
giant branch star in the proto-planetary nebula phase.

\end{abstract}

\keywords{stars: winds --- supernovae --- supernovae: individual (2002ic)}

\section{Introduction}
\label{sec:introduction}

Historically, the fundamental division of supernova (SN) types was
defined by the absence (Type~I) or presence (Type~II) of hydrogen in
the observed spectrum.  Later refinements distinguished Type~Ia
supernovae from other types of supernovae by the presence of strong
silicon absorption features in their spectra
\citep{wheeler90,filippenko97}. Type~Ia supernovae (SNe~Ia) are
generally accepted to result from 
the thermonuclear burning of a white dwarf in a binary
system, whereas all the other types of supernovae are believed to be
produced by the collapse of the stellar core, an event which leads to
the formation of a neutron star or black hole.

While interaction with circumstellar material (CSM) has been observed
for many core-collapse supernovae, the search for evidence of CSM
around Type~Ia~SNe has so far been unsuccessful. \citet{cumming96}
reported high resolution spectra of SN~1994D and found an upper limit
for the pre-explosion mass-loss rate of
$\dot{M}\sim1.5\times10^{-5}~M_\Sun~\mathrm{yr}^{-1}$ for an assumed
wind speed of $v_w = 10$~km~s$^{-1}$.
However, they also note that this limit allows most of the expected
range of mass-loss rates from symbiotic systems
($\frac{\dot{M}}{v_{10}} \lesssim 2\times10^{-5}~M_\Sun$~yr$^{-1}$).
On the other
hand, the surprisingly strong high-velocity Ca~II absorption and
associated high degree of linear polarization observed in SN~2001el 
by \citet{wang03a} and the high velocity features in SN~2003du by
\citet{gerardy04} have led these authors 
to suggest that the high velocity Ca feature could be
the result of the interaction between the supernova ejecta and a CSM
disk.  About 0.01~M$_\Sun$ of material is required in the disk, and
the spatial extent of the disk must be small to be consistent with the
absence of narrow emission lines at around optical maximum
\citep{cumming96}.  Due to the strength of the Ca~II feature in SN~2001el,
\citet{wang03a} speculated that the disk of SN~2001el may have been over-abundant in Ca~II. In contrast, \citet{gerardy04} found that a standard solar
abundance of Ca~II is sufficient to explain the observed feature in
SN~2003du~\citep{gerardy04}, for which the high-velocity Ca~II feature
is significantly weaker than in SN~2001el.

Supernova 2002ic~\citep{iauc8019b} is a very interesting event that shows both silicon
absorption~\citep{iauc8028} and hydrogen
emission~\citep{hamuy03b}. This SN is the first case for which there is
unambiguous evidence of the existence of circumstellar matter around a
SN~Ia and is therefore of great importance to the understanding of the
progenitor systems and explosion mechanisms of SNe~Ia.  By studying
the spectral polarimetry and the light curve of the H$\alpha$ line,
\citet{wang04} found the spatial extent of the hydrogen-rich 
material to be as large as 10$^{17}$~cm and distributed in a quite
asymmetric configuration, most likely in the form of a flattened
disk. The implied total mass of the hydrogen-rich CSM is a few solar
masses.  Similar conclusions were reached by \citet{deng04}.

In this paper, we present new photometry of SN~2002ic and discuss the
implications for the interaction of the ejecta and the CSM.
Sec.~\ref{sec:processing} presents our data processing procedure and
calibration for our photometry of SN~2002ic.  In
Sec.~\ref{sec:lightcurve}, we discuss the light curve of SN~2002ic and
the immediate implications from our data.  A more in-depth
investigation and qualitative modeling of the light curve of SN~2002ic
as an interaction of a SN~Ia with surrounding CSM is presented in
Sec.~\ref{sec:modeling}.  Our discussion in Sec.~\ref{sec:discussion}
presents our interpretations of the structure of the CSM surrounding
SN~2002ic.  Finally, in Sec.~\ref{sec:conclusions} we present some
intriguing possibilities for the progenitor system of SN~2002ic and
speculate on other possible SN~2002ic-like events.

\section{Data Processing for SN~2002ic}
\label{sec:processing}

\subsection{Data processing and Discovery}

We discovered SN~2002ic on images from the NEAT team~\citep{pravdo99}
taken on the Samuel Oschin 1.2-m telescope on Mt.~Palomar, California.
In preparation for searching, the images were transmitted from the
telescope to the High-Performance Storage System (HPSS) at the
National Energy Research and Scientific Computer Center (NERSC) in
Oakland, California via the
HPWREN~\citep{hpwren}\footnote{http://hpwren.ucsd.edu} and
ESnet~\citep{esnet}\footnote{http://www.es.net} networks.  These data were then
automatically processed and reduced on the NERSC Parallel Distributed
System Facility (PDSF) using software written at Lawrence Berkeley
National Laboratory by WMWV and the Supernova Cosmology Project.

The first-level processing of the NEAT images involved decompression
and conversion from the NEAT internal format used for transfer to the
standard astronomical FITS format, subtraction of the dark current for
these thermoelectrically cooled CCDs, and flat-fielding with sky flats
constructed from a sample of the images from the same night.  These
processed images were then loaded into an image database, and archival
copies were stored on HPSS.  The images were further 
processed to remove the sky background.  An object-finding algorithm
was used to locate and classify the stars and galaxies in the fields.
The stars were then matched and calibrated against the USNO~A1.0
{POSS-E} catalog~\citep{usnoa1} to derive a magnitude zeropoint for each
image.  There were typically a few hundred USNO~A1.0 stars in each
0.25~\sq\degr\ image.

The supernova was discovered by subtracting PSF-matched historical
NEAT images from new images, then automatically detecting residual
sources for subsequent human inspection (see \citet{wood-vasey04}).

\subsection{Photometry}

For analysis, we assembled all NEAT images, including later images
kindly taken at our request by the NEAT team.

Light curves were generated using aperture photometry scaled to the
effective seeing of each image.  A set of the 4 best-seeing
($<3$\arcsec) reference images was selected from among all NEAT
Palomar pre-explosion images from 2001 of SN~2002ic.  Multiple
reference images were chosen to better constrain any underlying galaxy
flux.  The differential flux in an aperture around SN~2002ic was
measured between each reference image and every other image of
SN~2002ic.  Aperture correction was performed to account for the
different seeing and pixel scales of the images.  The overall flux
ratio between each reference image and light-curve image was tracked
and normalized with respect to a primary reference image.  This
primary reference image was chosen from the reference images used for
the image subtraction on which SN~2002ic was originally discovered.
The flux differences calculated relative to each reference were
combined in a noise-weighted average for each image to yield an
average flux for the image.  As the observations were taken within a
span of less than one hour on each night, the results from the images
of a given night were averaged to produce a single light curve point
for that night.

The reference zeropoint calculated for the primary reference image
from the above USNO~A1.0 {POSS-E} calibration was used to set the
magnitudes for the rest of the measured fluxes.
Table~\ref{tab:2002ic_lightcurve} reports these magnitudes and
associated measurement uncertainties.  An overall systematic
uncertainty in the zeropoint calibration is not included in the listed
errors.  The USNO~A1.0 {POSS-E} catalog suffers from systematic
field-to-field errors of $\sim0.25$~magnitudes in the northern
hemisphere~\citep{usnoa1}.  The conversion of {POSS-E} magnitudes to
V-band magnitudes for a SN~Ia is relatively robust, as a SN~Ia near
maximum resembles a $\sim10,000$~K blackbody quite similar to Vega in
the wavelength range from $4,500$--$10,000$~\AA.  At late times,
the observations of~\citet{wang04} show that the smoothed spectrum of
SN~2002ic tracks that of Vega red-ward of 5,000~\AA.  We estimate
that, taken together, the calibration of our unfiltered observations
with these {POSS-E} magnitudes and the subsequent comparison with V-band
magnitudes
are susceptible to a $0.4$~magnitude systematic uncertainty.  Any such
systematic effect is constant for all data points and stems directly
from the magnitude calibration of the primary reference.

However, the observed NEAT {POSS-E} magnitudes show agreement with the
V-band magnitudes of \citet{hamuy03b} and the V-band magnitudes
obtained from integrating the spectrophotometry of \cite{wang04} to
significantly better than any $0.4$~magnitude systematic uncertainty
estimate.  This synthesized VLT photometry is presented in
Table~\ref{tab:2002ic_VLT}.  Comparing the photometry of SN~2002ic and
nearby reference star with a similar color ($B-R=0.3$), we find agreement
between the VLT V-band acquisition camera images and the NEAT images
to within $\pm 0.05$~magnitudes.  Given this good agreement, it appears
that our POSS-E-calibrated magnitudes for SN~2002ic can be used effectively as V-band
photometry points.

Mario Hamuy was kind enough to share his BVI secondary standard stars
from the field of SN~2002ic.  We attempted to use these stars to
calculate the color correction from our {POSS-E} magnitudes to V-band,
but our analysis predicted an adjustment of up to $+0.4$~magnitudes.
This was inconsistent with the
good agreement with the VLT magnitudes (calculated correction =
$+0.1$~mag) at late times and with the Hamuy V-band points after
maximum ($+0.4$~mag).  This disagreement is not fully understood.  We
note, however, that the colors of the secondary standard stars did not
extend far enough to the blue to cover the majority of the color range
of the supernova during our observations (a common problem when
observing hot objects such as supernovae).  In addition, as there is no 
color information from before maximum light, it is possible that
SN~2002ic does not follow the color evolution of a typical SN~Ia.

\begin{figure}
\plotone{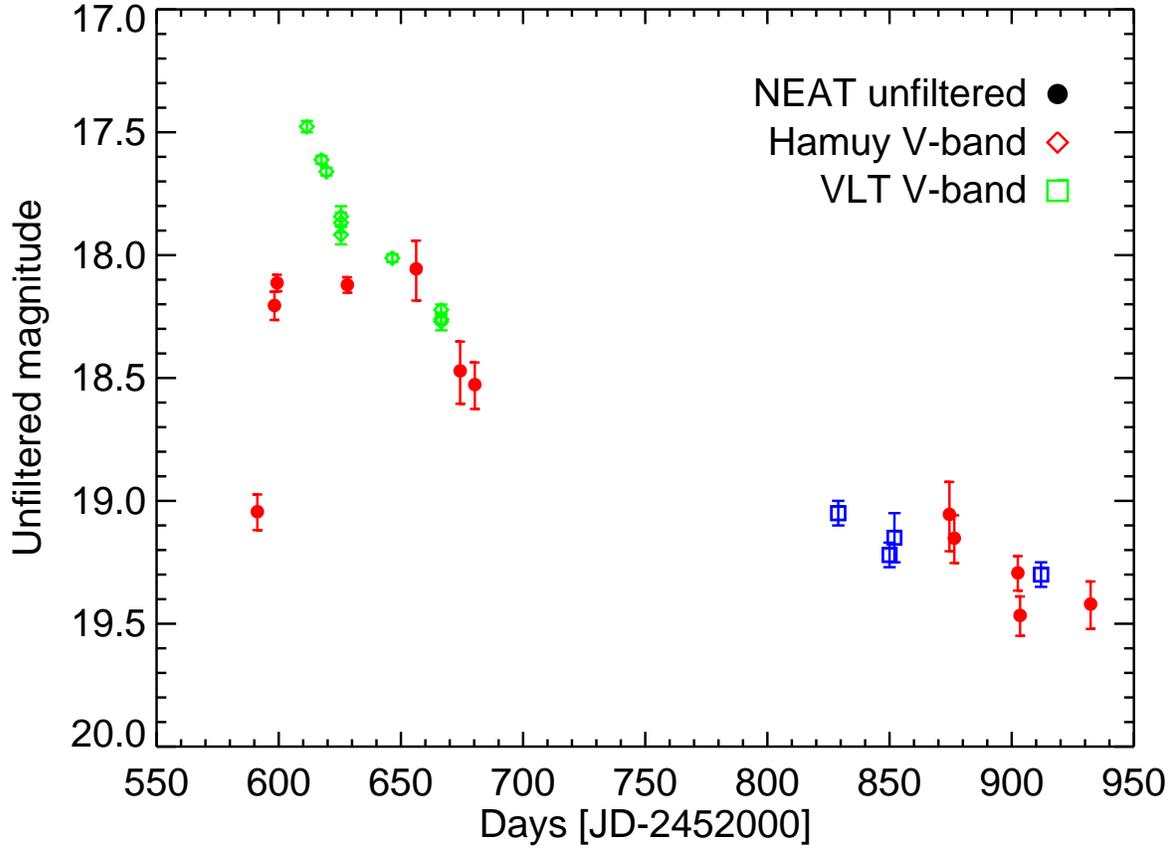}
\caption{The unfiltered optical light curve of SN~2002ic as observed by NEAT with
the Palomar 1.2-m and Haleakala 1.2-m telescopes (see
Table~\ref{tab:2002ic_lightcurve}).  The magnitudes have been
calibrated against the USNO-A1.0 {POSS-E} stars in the surrounding
field.  No color correction has been applied.  Also shown 
are the observed V-band magnitudes from~\citet{hamuy03b} and
V-band magnitudes from the spectrophotometry of~\citet{wang04}.}
\label{fig:2002ic_VLT}
\end{figure}

Our newly reported pre-maximum photometry points (see
Table~\ref{tab:2002ic_lightcurve} and Fig.~\ref{fig:2002ic_VLT}) are
invaluable for disentangling the SN and CSM components, which we now
proceed to do.

\section{Light Curve of SN~2002ic}
\label{sec:lightcurve}

The light curve of SN~2002ic is noticeably different from that of a
normal SN~Ia, as can be seen in Fig.~\ref{fig:2002ic_template}, and as
was first noted by \citet{iauc8151}.  The detection of hydrogen
emission lines in the spectra of SN~2002ic in combination with the
slow decay of the light curve is seen as evidence for interaction of
the SN ejecta and radiation with a hydrogen-rich
CSM~\citep{hamuy03b,wang04}.  The profile of the hydrogen emission
line and the flat light curves can be understood in the context of
Type~IIn supernovae as discussed in \citet{chugai02}, \citet{chugai04}, and references therein. 

The data presented here show that the
slow decay has continued $\sim320$~days after maximum at a rate of
$\sim0.004$~mag/day, a rate that is significantly slower than the
$0.01$~mag/day decay rate expected from Co$^{56}$ decay (also see
\citet{deng04}).  In addition, our early-time points show that the
light curve of SN~2002ic was consistent with a pure SN~Ia early in its
evolution.  This implies that there was 
a significant time delay between the explosion and development
of substantial radiation from the CSM interaction, possibly due to a 
a physical gap between the
progenitor explosion and the beginning of the CSM.  After maximum, we
note the existence of a second bump in the light curve, which is put in
clear relief by our photometry data on JD~$2452628.6$.  We interpret
this second bump as evidence for further structure in the CSM.

\section{Decomposition of SN Ia and CSM components}
\label{sec:decomposition}

\begin{figure}
\plotone{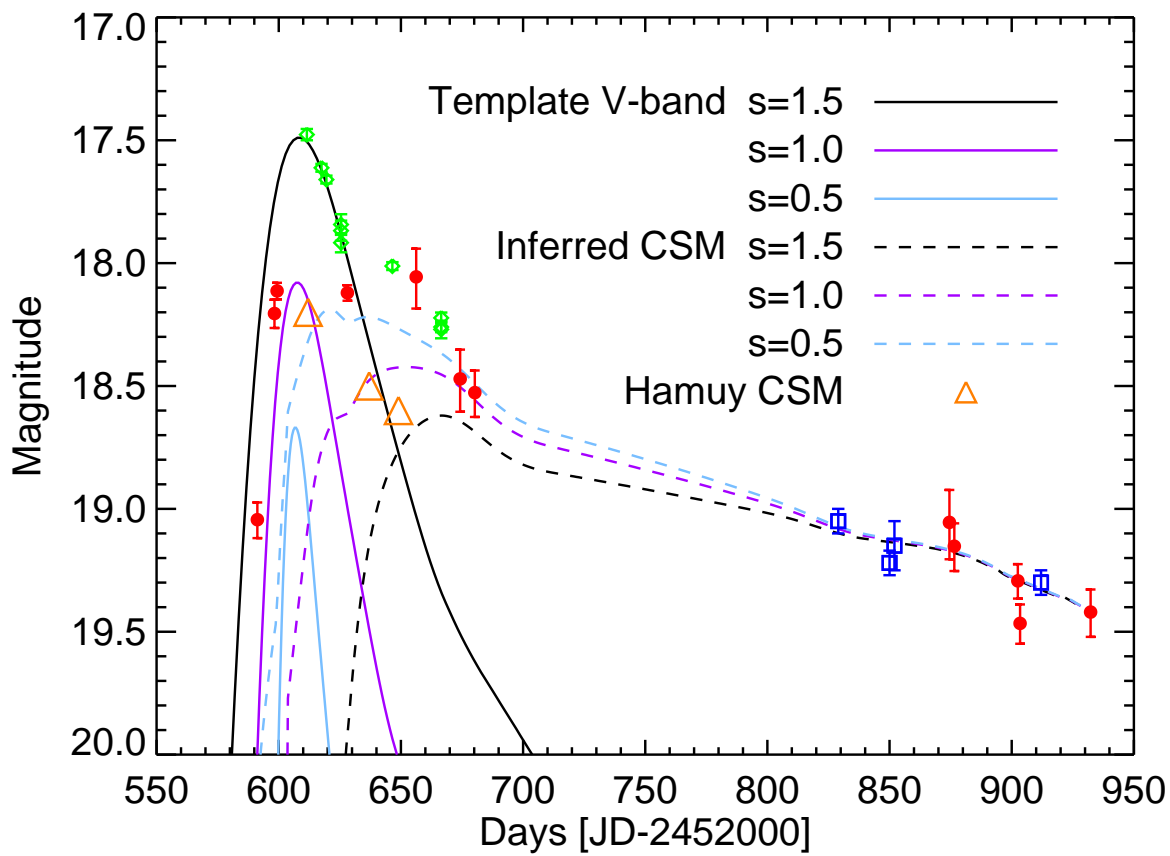}
\caption{
A template SN~Ia V-band light curve (solid lines -- stretch decreases from top to bottom line) shown for comparison 
with the photometric observations at several
stretch values, $s$, where the magnitude-stretch relation 
$\Delta m = 1.18~(1-s)$
has been applied.  The difference between the observed
photometry points and the template fit has been smoothed over a $50$-day window (dashed lines).  Note that an assumption of no CSM
contribution in the first $15$ days after maximum light 
(i.e. $s=1.5$)
is in conflict
with the spectroscopic measurements of \citet{hamuy03b} (open triangles--no error bars available).  }
\label{fig:2002ic_template}
\end{figure}

\citet{hamuy03b} performed a spectroscopic decomposition of the 
underlying supernova and ejecta-CSM interaction components.  We
perform here an analogous photometric decomposition.  To decompose the
observed light curve into the contributions from the SN
material and the shock-heated CSM, we first consider a range of
light curve stretch values~\citep{perlmutter97b}, using
the magnitude-stretch relation, $\Delta m =1.18~(1-s)$~\citep{knop03}, 
applied to the normal SN~Ia template light
curve of \citet{goldhaber01}; we consider the remaining flux as being
due to the SN~eject-CSM interaction (see
Fig.~\ref{fig:2002ic_template}).
At early times, the inferred contribution of the CSM is dependent on
the stretch of the template chosen, but at later times the CSM
component completely dominates for any value of the stretch parameter.  It is not possible to disentangle the
contribution of the CSM from that of the SN at maximum light, although
a normal SN~Ia at the redshift of SN~2002ic,
$z=0.0666$~\citep{hamuy03b}, corresponding to a distance modulus of
$37.44$ for an $H_0=72$~km/s/Mpc~\citep{freedman01}, would only generate about half of
the flux observed for SN~2002ic at maximum.
\citet{hamuy03b} find that SN~2002ic resembles SN~1991T
spectroscopically and note that SN~1991T/SN~1999aa-like events are
brighter a month after maximum light than explainable by the standard
stretch relation.  A SN~1991T-like event (stretch$=1.126$, $\Delta
m=0.15$, based on the template used in \citet{knop03} (A.J. Conley
2004, private communication)), would lie near the stretch$=1$ line of
Fig.~\ref{fig:2002ic_template}.  The light curve of SN~2002ic for the
first 50~days is thus much too luminous to be due entirely to a
91T-like supernova.  In addition, the spectroscopically-inferred
CSM-interaction contribution of \citet{hamuy03b} 
(open triangles in Fig.~\ref{fig:2002ic_template}) limits the SN
contribution at maximum to that expected from a normal SN~Ia.  After
50~days, SN~2002ic exhibits even more significant non-SN~Ia-like
behavior.

We next use the formalism of \citet{chevalier94} to fit a simple
interacting SN ejecta-CSM model to the observed data.  While 
\citet{chevalier94} focus on SNe~II, their formalism is generally
applicable to SNe ejecta interacting with a surrounding CSM.
We simultaneously fit the SN~Ia flux and the luminosity from the SN
ejecta-CSM interaction.  Our analysis allows us to infer the integrated radial
density distribution of the CSM surrounding SN~2002ic.  

\section{Simple Scaling of the SN Ejecta-CSM Interaction}
\label{sec:modeling}

Following the hydrodynamic models of \citet{chevalier94}, we assume a
power-law supernova ejecta density of
\begin{equation}
\rho_\mathrm{SN} \propto t^{n-3} r^{-n}
\label{eq:ejecta_density}
\end{equation}
where $t$ is the time since explosion, $r$ is the radius of the ejecta,
and $n$ is the power-law index of a radial fall-off in the ejecta density.
\cite{chevalier01} note that for SNe~Ia an exponential ejecta profile 
is perhaps preferred.  However, this profile does not yield an analytical
solution and so, for the moment, we proceed assuming a power-law profile.
In Sec.~\ref{sec:discussion} we explore the ramifications of an 
exponential ejecta profile.

\cite{chevalier94} give the time evolution of the shock-front radius, $R_s$, as
\begin{equation}
R_s = \left [ \frac{2}{(n-3)(n-4)} \frac{4\pi v_w}{\dot{M}} A \right]^{1/(n-2)} t^{(n-3)/(n-2)},
\label{eq:shock_radius}
\end{equation}
where $v_w$ is the velocity of the pre-explosion stellar wind,
$\dot{M}$ is the pre-explosion mass-loss rate, and
$A$ is a constant in the appropriate units for the given
power-law index $n$.

Taking the parameters in the square brackets as fixed constants, we can
calculate the shock velocity, $v_s$, as
\begin{equation}
v_s = \left [ \frac{2}{(n-3)(n-4)} \frac{4\pi v_w}{\dot{M}} A \right]^{1/(n-2)} \left(\frac{n-3}{n-2}\right) t^{-1/(n-2)}.
\label{eq:shock_velocity}
\end{equation}
Thus the shock velocity goes as 
\begin{equation}
v_s \propto t^{-\alpha},
\label{eq:vs_time}
\end{equation}
where
\begin{equation}
\alpha = \frac{1}{n-2}.
\label{eq:vs_time_alpha}
\end{equation}

We assume that the luminosity of the ejecta-CSM interaction is fed by
the energy imparted at the shock front and view the unshocked wind as
crossing the shock front with a velocity of $v_s + v_w \approx v_s$.
As the wind particles cross the shock front, they are thermalized and
their crossing kinetic energy, $\mathrm{K.E.}=\onehalf \rho_w v_s^2 dV$, is
converted to thermal energy.  Putting this in terms of the mass-loss
rate, $\dot{M}$, we can express the CSM density as
\begin{equation}
\rho_w = \frac{\dot{M}}{4\pi R_s^2 v_w},
\end{equation}
and we can calculate the energy available to be converted to luminosity, $L$, as
\begin{equation}
L = \alpha(\lambda,t) \frac{d}{dt}\mathrm{K.E.} = \alpha(\lambda,t) \frac{1}{2} \frac{\dot{M}}{4\pi R_s^2 v_w} v_s^2 dV 
= \alpha(\lambda,t) \frac{1}{2} \frac{\dot{M}}{4\pi R_s^2 v_w} v_s^2 v_s 4\pi R_s^2.
\end{equation}
The luminosity dependence on $R_s$ drops out and we have
\begin{equation}
L = \alpha(\lambda,t) \frac{d}{dt} \mathrm{K.E.} = \alpha(\lambda,t) \frac{1}{2} \frac{\dot{M}}{v_w} v_s^3.
\end{equation}
A key missing ingredient is a more detailed modeling of the kinetic
energy to optical luminosity conversion term, $\alpha(\lambda, t)$.
We note that the available kinetic energy is on the order of $1.6\times
10^{44}$~erg~s$^{-1}$ for $\dot{M} = 10^{-5}~M_\sun$~yr$^{-1}$, $v_s =
10^4$~km~s$^{-1}$, and $v_w = 10$~km~s$^{-1}$.  This implies a
conversion efficiency from shock interaction K.E. to luminosity of
50\%, given the luminosity, $1.6\times10^{44}$~erg~s$^{-1}$, of
SN~2002ic and the typical luminosity of a SN~Ia near maximum of
$0.8\times10^{44}$~erg~s$^{-1}$.  Assuming this constant conversion
produces reasonable agreement with the data, so we proceed with this
simple assumption.  Using Eq.~\ref{eq:vs_time} to give the time
dependence of $v_s$, we obtain the time dependence of the luminosity,
\begin{equation}
L \propto v_s^3 \propto t^{-3\alpha},
\end{equation}
which can be expressed in magnitude units as
\begin{equation}
m_\mathrm{ejecta-CSM} = C - \frac{5}{2} \log_{10}{t^{-3\alpha}} = C + \frac{15}{2}\alpha \log_{10}{t},
\end{equation}
where $C$ is a constant that incorporates $\dot{M}$,
$\rho_{\mathrm{SN}}$, $v_w$, $n$, and the appropriate units for those
parameters.
The difference in magnitude between two times, $t_1$ and $t_2$, then becomes
\begin{equation}
m_{t_2} - m_{t_1} = \frac{15}{2} \alpha \log_{10}{\frac{t_2}{t_1}}.
\end{equation}

We obtain a date of B-maximum for the supernova component of
$2452606$~JD from our SN~Ia-light curve analysis.
Our fit yields an $\alpha=0.16 \Rightarrow n = 8.5$
for any fixed $\dot{M}$ and $v_w$.  This $n$ is squarely in the range
of values suggested by \citet{chevalier94} as being typical for SN
ejecta.  While \citet{chevalier94} is framed in the context of 
SNe~II, their formalism applies to any SN explosion into a surrounding medium
whose ejecta density profile is described by their analytic model.

The interaction scaling relations presented above are useful for
decomposing the interaction and supernova contributions to the total
light curve of SN~2002ic. This simple, analytic description 
approximates our data reasonably well.
However, more sophisticated theoretical calculations,
which are beyond the scope of this paper,
are necessary
to more quantitatively derive the detailed physical parameters of the
SN ejecta and the CSM (see \citet{chugai04}).

\section{Discussion}
\label{sec:discussion}

\subsection{Inferred CSM structure and Progenitor Mass-Loss History}

\begin{figure}
\plotone{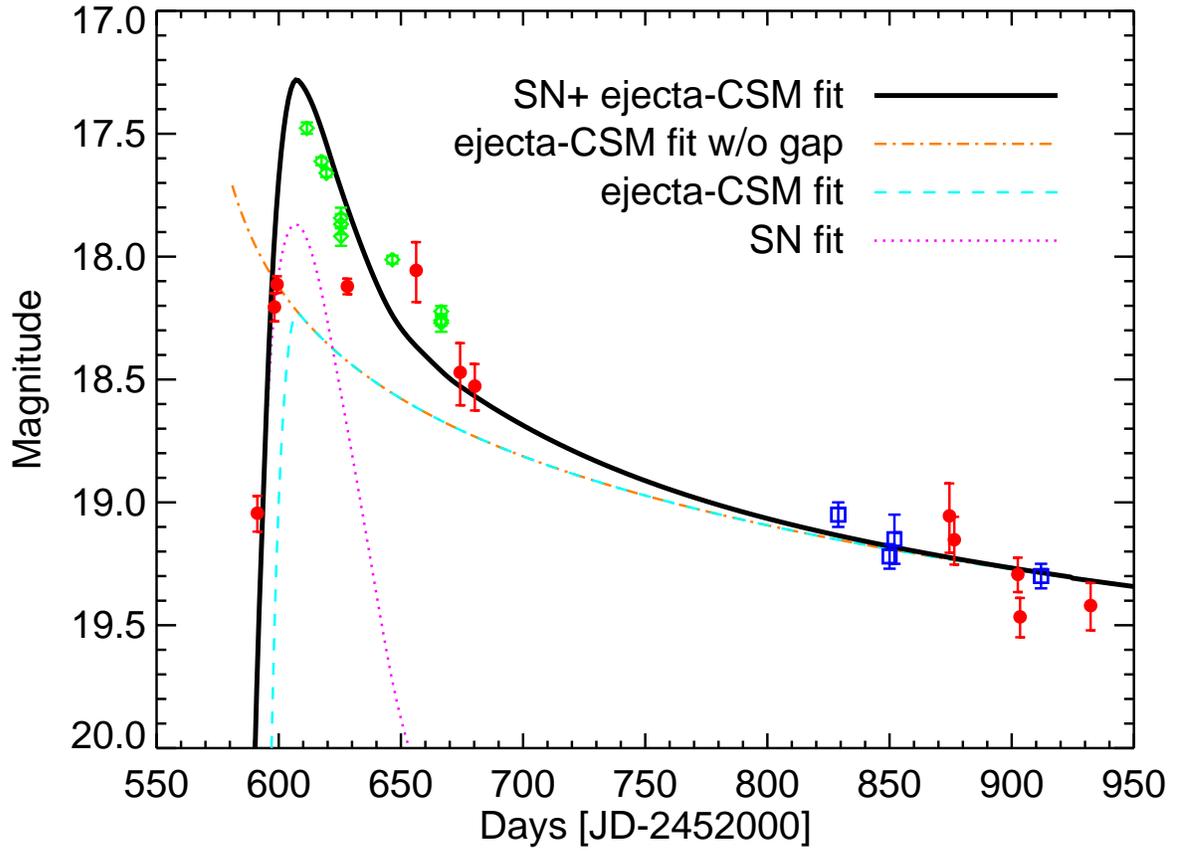}
\caption{
The observed 
photometry compared with
the SN + power-law ejecta-CSM model described in Sec.~\ref{sec:modeling}.
}
\label{fig:2002ic_template_csm_fit}
\end{figure}

We can match the inferred SN ejecta-CSM component of \citet{hamuy03b}
with the interaction model described above and reproduce the light
curve near maximum light by adding the flux from a normal SN~Ia.
Fig.~\ref{fig:2002ic_template_csm_fit} shows our model fit in
comparison with the observed light curve of SN~2002ic.  Note that our
model does not match the observed bump at 40~days after maximum.

\citet{hamuy03b} note a similar disagreement, but the data we present here 
show that this region is clearly a second bump rather than just a very slow
decline.  This discrepancy could be explained by a change in the
density of the circumstellar medium due to a change in the progenitor
mass-loss evolution at that point.  In fact, our simple fit is too bright
before the bump and too dim during the bump, implying more structure
in the underlying CSM than accounted for in our model.  
Any clumpiness in the progenitor wind would have to be on the largest
angular scale to create such a bump and would not explain the new
decline rate shown by our observations to extend out to late time .
We find that our data are consistent with a model comprising three CSM
density regions: (i) an evacuated region out to $20v_s$~days; (ii)
CSM material at a nominal density ($\rho\propto r^{-2}$) out to
$\sim100v_s$~days; and (iii) an increase in CSM density at
$\sim100v_s$~days, with a subsequent $r^{-2}$ fall-off extending
through the $800v_s$~days covered by our observations.
This model agrees well with the light curve of SN~2002ic, but, as it
involves too many parameters to result in a unique fit using only the
photometric data, we do not show it here.

Our data, particularly the pre-maximum observations,
provide key constraints on the nature of the progenitor system
of SN~2002ic.  In the context of our model, a mass-loss gradient of some form
is required by our early data points.
As a computational convenience, our model assumes that the transition
to a nominal circumstellar density is described by
$\sin(\frac{t}{20~\mathrm{days}})$.  If the mass-loss rate had been
constant until just prior to the explosion, then the $t^{-3\alpha}$
model light curve would continue to curve upward and significantly
violate our first data point at the $7\sigma$ level (as shown by the
line extended from the ejecta-CSM component in
Fig.~\ref{fig:2002ic_template_csm_fit}).  If the conversion of kinetic
energy to luminosity is immediate and roughly constant in time, as
assumed in our model, we would conclude that a low-density region must
have existed between the star and \mbox{$20$~days$\cdot v_s$}
out from the star.  For example, as a stellar system transitions from
an AGB star to a proto-planetary nebula (PPN), it changes from
emitting a denser, cooler wind, to a hotter, less dense
wind~\citep{kwok93}.  This hot wind pushes the older wind out farther
and creates a sharp density gradient and possible clumping near the
interface between the cool and hot winds~\citep{young92}.  This
overall structure is similar to that which we infer from our modeling
of SN~2002ic.  Assuming a SN ejecta speed of $v_s=30,000$~km~s$^{-1}$
and a progenitor star hot wind speed of
$v_w=100$~km~s$^{-1}$~\citep{young92,herpin02}, we conclude that the
hot wind must have begun just $\sim15$~years prior to the SN
explosion.  Alternatively, there is also the possibility that the
conversion from kinetic energy to optical luminosity is for some
reason significantly less efficient at very early times.

It is interesting to note that the observed light curve decline rate
of SN~2002ic after $40$~days past maximum light is apparently constant
during these observations.  Spectroscopic study
\citep{wang04} shows the highest observed velocity of the ejecta
to be around $11000$~km~s$^{-1}$ at day~$200$ after maximum light.
If we assume a constant expansion rate, these observations of continuing
emission through $\sim320$~days after maximum provide a lower limit of
$\sim3\times10^{16}$~cm for the spatial extent of the CSM.  Compared to a
nominal pre-explosion stellar wind speed of $10$~km~s$^{-1}$, the ejecta is
moving $\sim1000$ times more rapidly and thus has overtaken the progenitor
wind from the past $\sim800$~years.  The overall smoothness of the
late-time light curve shows the radial density profile of the CSM to
be similarly smooth and thus implies a fairly uniform mass-loss rate
between $100$--$800$ years prior to the SN explosion.

We take the lack of enhanced flux at early times and the bump after
maximum light as evidence for a gap between the SN progenitor and the
dense CSM as well as a significant further change in the mass-loss of
the progenitor system $\sim100$~years prior to the SN explosion.

\subsection{Reinterpretation of Past SNe~IIn}

These new results prompt a reexamination of supernovae previously
classified as Type~IIn, specifically SN~1988Z \citep{iauc4691,stathakis91},
SN~1997cy~\citep{iauc6706,turatto00,germany00}, and
SN~1999E~\citep{iauc7091,siloti00,rigon03}. These supernovae bear
striking similarities in their light curves and their late-time
spectra to SN~2002ic.  However, SN~2002ic is the only one of these
supernovae to have been observed early in its evolution.  If SN~2002ic
had been observed at the later times typical of the observations of
these Type IIn SNe, it would not have been identified as a Type~Ia.
It is interesting to note that \citet{chugai94} found from models of
light curves of SN~1988Z that the mass of the SN~1988Z supernova
ejecta is on the order of 1~$M_\sun$, which is consistent with a
SN~Ia.

\begin{figure}
\plotone{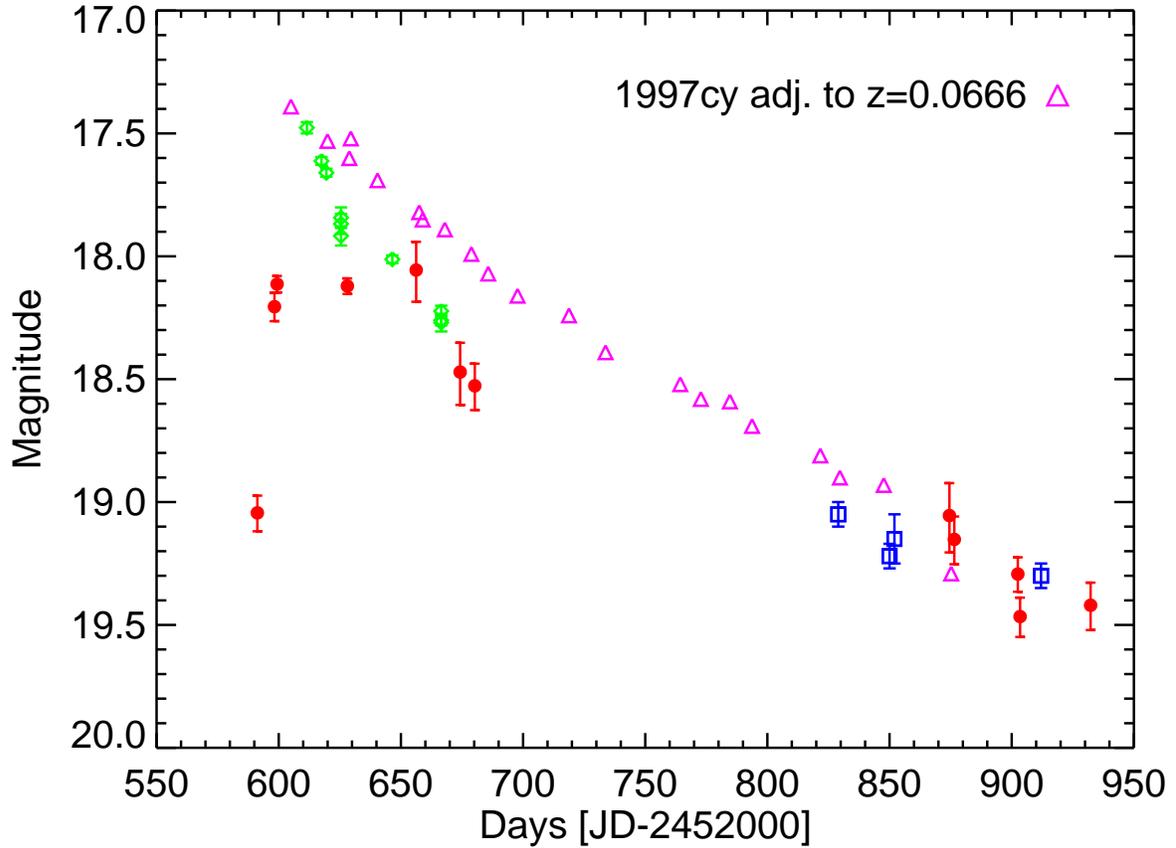}
\caption{The NEAT unfiltered and Hamuy V-band observations of SN~2002ic
compared to the K-corrected V-band observations of SN~1997cy from
\citet{germany00}.  No date of maximum or magnitude uncertainties are
available for SN~1997cy.  Here the maximum observed magnitude for SN~1997cy
has been adjusted to the redshift of 2002ic, z=0.0666~\citep{hamuy03b}, 
and the date of the first light curve point of SN~1997cy has been set to the 
date of maximum for SN~2002ic from our V-band fit.}
\label{fig:2002ic_1997cy}
\end{figure}

\begin{figure}
\plotone{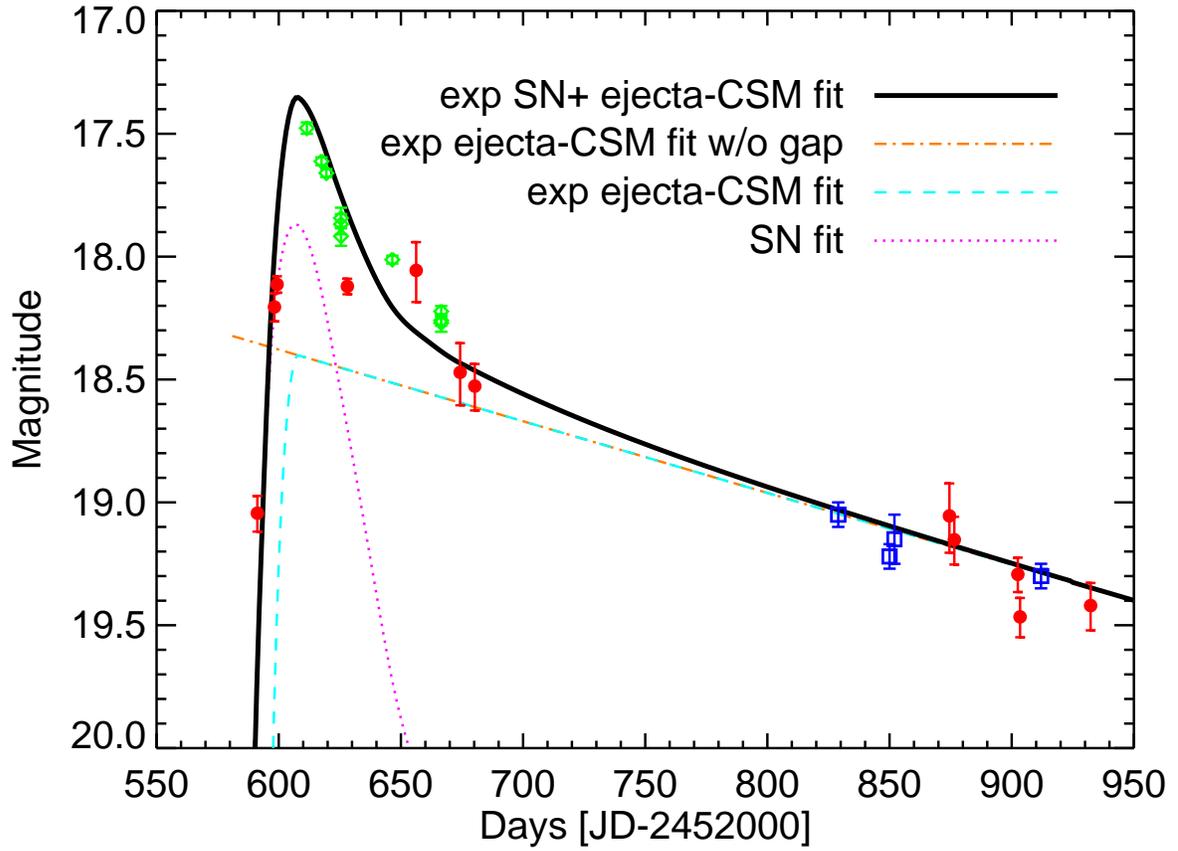}
\caption{
The observed 
photometry compared with
the SN + exponential ejecta-CSM model described in Sec.~\ref{sec:discussion}.
}
\label{fig:2002ic_template_csm_exp_fit}
\end{figure}

\begin{figure}
\plotone{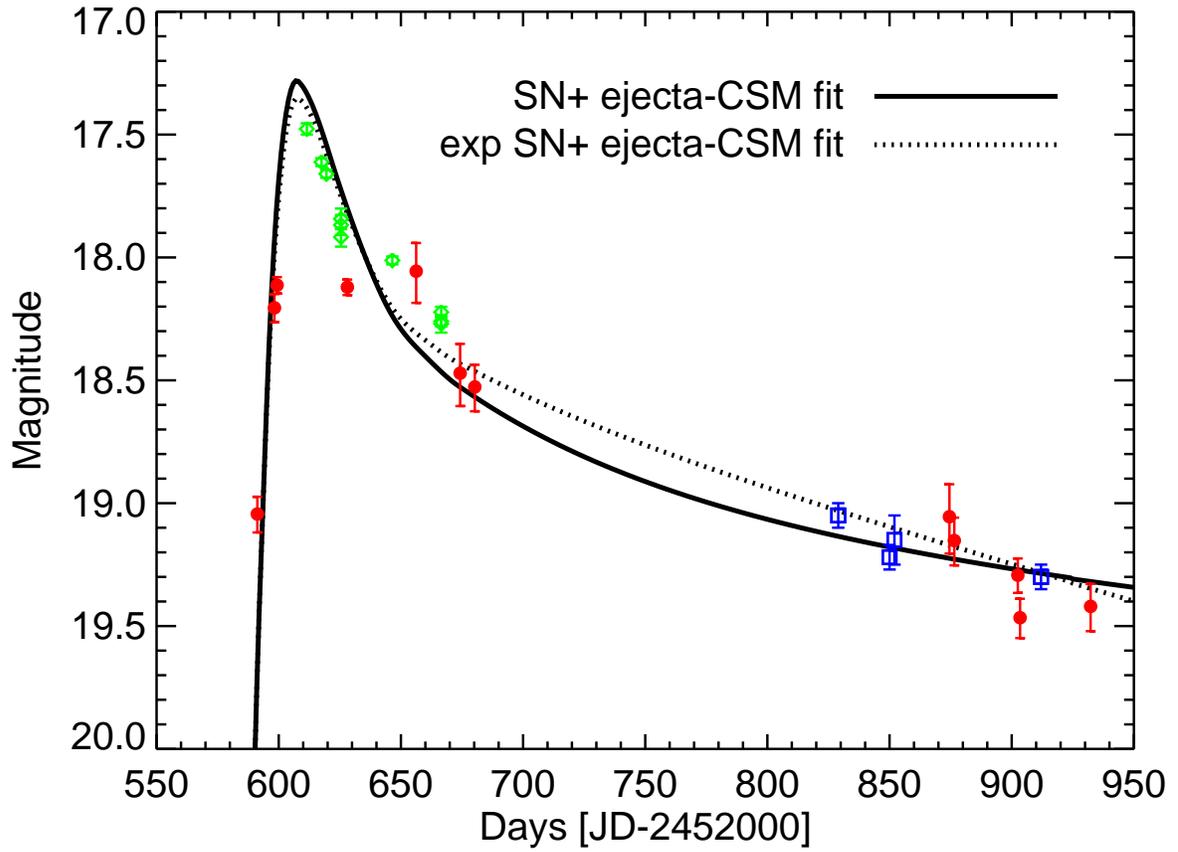}
\caption{
A comparison of fits with power-law (solid) and exponential (dotted) SN ejecta density profiles.
}
\label{fig:2002ic_template_csm_both_fit}
\end{figure}

We next explore the possibility that SN~1997cy and SN~1999E (a close
parallel to SN~1997cy) may have been systems like SN~2002ic.
\citet{hamuy03b} found that available spectra of SN~1997cy were 
very similar to post-maximum spectra of SN~2002ic.  We complement this
spectroscopic similarity with a comparison of the photometric behavior
of SN~1997cy and SN~2002ic.  As shown in Fig.~\ref{fig:2002ic_1997cy},
the late-time behavior of both SNe appear remarkably similar with both
SNe fading by $\sim2.5$~magnitudes 8 months after their respective
discoveries.  The luminosity decay rate of the ejecta-CSM interaction
is directly related to the assumed functional form for the ejecta
density and the mass-loss rate (Eq.~\ref{eq:ejecta_density}).  The
observed late-time light curves of SN~1997cy and SN~1999E clearly
follow a linear magnitude decay with time, which implies an
exponential flux vs. time dependence: $m \propto t \Rightarrow
\mathrm{flux} \propto e^{Ct}$.  If the ejecta density followed an
exponential rather than a power-law decay, the magnitude would
similarly follow a linear magnitude-time decay.
Fig.~\ref{fig:2002ic_template_csm_exp_fit} shows a fit to the light
curve of SN~2002ic using the framework of Sec.~\ref{sec:modeling} 
but using an exponential SN-ejecta density profile.
\citet{chevalier01} suggest that SNe~Ia follow exponential
ejecta profiles~\citep{dwarkadas98} while core-collapse SNe follow
power-law decays~\citep{chevalier89,matzner99}.  Thus, if SN~1997cy
and SN~1999E had been core-collapse events, they would have been
expected to show power-law declines.  Instead, their decline behavior
lends further credence to the idea that they were SN~Ia events rather
than core-collapse SNe.  Although we modeled the light curve of
SN~2002ic using a power-law ejecta profile, SN~2002ic was not observed
between 100 and 200 days after explosion, so its decay behavior during
that time is not well constrained.  Its late-time light curve is
consistent with the linear magnitude behavior of SN~1997cy.  We fit
such a profile to our data (see
Fig.~\ref{fig:2002ic_template_csm_exp_fit}~\&~\ref{fig:2002ic_template_csm_both_fit})
and arrive at an exponential fit to the flux of the form $e^{-0.003
t}$ where $t$ is measured in days.  As the solution for the SN-ejecta
interaction is not analytic, we cannot immediately relate the
exponential decay parameter to any particular property of the SN
ejecta.  Taken together in the context of the \citet{chevalier94}
model, SN~2002ic, SN~1997cy, and SN~1999E lend support to numerical
simulations of the density profiles of SNe~Ia explosions.

If we take the time of maximum for SN~1997cy to be the earliest light curve
point from \citet{germany00} and shift the magnitudes from the
redshift of SN~1997cy, $z=0.063$~\citep{germany00}, to the redshift of
SN~2002ic ($z=0.0666$), we find that the luminosity of both SNe agree
remarkably well.  This further supports that hypothesis that SN~1997cy
and SN~2002ic are related events.  However, we note that the explosion
date of SN~1997cy is uncertain and may have been 2--3 months prior to
the discovery date~\citep{germany00,turatto00}.

Fig.~\ref{fig:2002ic_template_csm_fit}~\&~\ref{fig:2002ic_template_csm_exp_fit}
show that neither a power-law nor an exponential model allow for a
significant ejecta-CSM contribution before maximum light.  In each
figure, the ``[exp] ejecta-CSM fit w/o gap'' line shows how the SN
ejecta-CSM interaction would continue if the density profile remained
the same.  Both lines significantly disagree with the earliest light
curve point.  This is consistent with our earlier conclusion that the light
curve is dominated by the SN until near maximum light.

\subsection{Relation to Proto-Planetary Nebulae?}

The massive CSM and spatial extent inferred for SN~2002ic
are surprisingly similar to certain
PPNe and the atmospheres of very late red giant
stars evolving to PPNe. Such structures are normally short-lived (less
than or on the order of $1000$~years).  The polarization seen by
\citet{wang04} suggests the presence of a disk-like structure
surrounding SN~2002ic.  Furthermore, the H$\alpha$ luminosity and mass
and size estimates suggest a clumpy medium.  Combined with the
evidence presented here for a possible transition region between a
slow and fast wind, we are left with an object very similar to
observed PPNe.  We encourage more detailed radiative hydrodynamic
modeling of SNe~Ia in a surrounding medium as our data provide
valuable constraints on this important early-time phase.

Of particular interest are bi-polar PPNe where a
WD companion emits a fast wind that shapes the AGB star wind while
simultaneously accreting~\citep{soker00} mass from the AGB star.

Typical thermonuclear supernovae are believed to have accretion 
time scales of $10^{7}$~years, yet several SNe Ia (SN~2002ic and
possibly SN~1988Z, SN~1997cy, and SN~1999E) out of several hundred
have been observed to show evidence for significant CSM.  If the
presence of a detectable CSM is taken as evidence that these SNe
exploded within a particular $\sim1000$~year-period in their
respective evolution, such as the PPN phase, this coincidence would
imply a factor of $\sim100$ enhancement ($10^7$~years~$ /
1000$~years~$/100$) in the supernova explosion rate during this
period.  Thus we suggest that it is not a coincidence that the
supernova explosion is triggered during this phase.

\section{Conclusions}
\label{sec:conclusions}

The supernova 2002ic exhibits the light curve behavior and hydrogen
emission of a Type~IIn supernova after maximum but was
spectroscopically identified as a Type~Ia supernova near maximum
light.  The additional emission is attributed to a contribution from
surrounding CSM.  This emission remains quite significant
$\sim$11~months after the explosion.  The discovery of dense CSM 
surrounding a Type~Ia supernova strongly
favors the binary nature of Type~Ia progenitor systems to explain the
simultaneous presence of at least one degenerate object and
substantial material presumably ejected by a significant stellar wind.
However, it is as yet unclear whether the available data for SN~2002ic 
can prove or disprove either the single- or the double-degenerate scenario,
although the inferred resemblance to PPN systems is suggestive.
The early-time light curve data presented in this paper strongly
suggest the existence of a
$\sim15~(\frac{v_w}{10~\mathrm{km/s}})^{-1}$~year gap between the
exploding object and the surrounding CSM.
Our discovery and early- through
late-time photometric followup of SN~2002ic suggests a
reinterpretation of some Type~IIn events as Type~Ia thermonuclear
explosions shrouded by a substantial layer of circumstellar material.

\section{Acknowledgments}
\label{sec:acknowledgments}

We would like to acknowledge our fruitful collaboration with the NEAT
group at the Jet Propulsion Laboratory, operated under contract NAS7-030001 with the National Aeronautics and Space Administration, which provided the images for our supernova work.  
We thank Mario Hamuy for sharing his BVI photometry of stars in the
field of SN~2002ic. 
We are grateful to Nikolai Chugai for helpful comments.
This research used resources of the National Energy Research Scientific
Computing Center, which is supported by the Office of Science of the U.S.
Department of Energy under Contract No. DE-AC03-76SF00098. We would like
to thank them for making available the significant computational 
resources required for our supernova search.  In addition, we thank
NERSC for a generous allocation of computing time on the IBM SP
RS/6000 used by Peter Nugent and Rollin Thomas to
reconstruct the response curve for the NEAT detector on our behalf.
HPWREN is operated by the University of California, San Diego under
NSF Grant Number ANI-0087344.
This work was supported in part by the Director, Office of Science,
Office of High Energy and Nuclear Physics, of the US Department of
Energy under Contract DE-AC03-76SF000098.
WMWV was supported in part by a National Science Foundation Graduate
Research Fellowship.

We thank the anonymous referee for helpful and detailed comments
that improved the scientific clarity of this manuscript.

\begin{deluxetable}{lrllr}
\tablewidth{0pc}
\tablecaption{The unfiltered magnitudes for SN~2002ic as observed by the
NEAT telescopes and shown in Fig.~\ref{fig:2002ic_VLT}.  The left
brackets ([) denote limiting magnitudes at a signal-to-noise of 3.  A
systematic uncertainty of $0.4$~magnitudes in the overall calibration
is not included in the tabulated uncertainties.  (See
Sec.~\ref{sec:processing} for further discussion of our calibration). }
\tablehead{
\colhead{JD - 2452000} & \colhead{E Mag} & \multicolumn{2}{c}{E Mag} & \colhead{Telescope} \\
\colhead{}             & \colhead{}      & \multicolumn{2}{c}{Uncertainty} & \colhead{}    
}
\startdata
    195.4999 &  [  20.52  &            &            &        Palomar 1.2-m \\
    224.2479 &  [  20.44  &            &            &        Palomar 1.2-m \\
    250.2492 &  [  21.01  &            &            &        Palomar 1.2-m \\
    577.4982 &  [  20.29  &            &            &      Haleakala 1.2-m \\
    591.2465 &     19.04  &   $-$  0.07  &   $+$  0.07  &        Palomar 1.2-m \\
    598.2519 &     18.20  &   $-$  0.06  &   $+$  0.06  &        Palomar 1.2-m \\
    599.3306 &     18.11  &   $-$  0.03  &   $+$  0.03  &        Palomar 1.2-m \\
    628.0956 &     18.12  &   $-$  0.03  &   $+$  0.03  &        Palomar 1.2-m \\
    656.2508 &     18.06  &   $-$  0.13  &   $+$  0.12  &      Haleakala 1.2-m \\
    674.2524 &     18.47  &   $-$  0.13  &   $+$  0.12  &      Haleakala 1.2-m \\
    680.2519 &     18.53  &   $-$  0.10  &   $+$  0.09  &      Haleakala 1.2-m \\
    849.5003 &  [  18.88  &            &            &      Haleakala 1.2-m \\
    853.4994 &  [  18.54  &            &            &      Haleakala 1.2-m \\
    855.4963 &  [  19.32  &            &            &      Haleakala 1.2-m \\
    858.4986 &  [  19.23  &            &            &      Haleakala 1.2-m \\
    860.4992 &  [  18.74  &            &            &      Haleakala 1.2-m \\
    864.5017 &  [  17.17  &            &            &      Haleakala 1.2-m \\
    874.4982 &     19.05  &   $-$  0.15  &   $+$  0.13  &      Haleakala 1.2-m \\
    876.4998 &     19.15  &   $-$  0.10  &   $+$  0.09  &      Haleakala 1.2-m \\
    902.4989 &     19.29  &   $-$  0.07  &   $+$  0.07  &        Palomar 1.2-m \\
    903.4138 &     19.47  &   $-$  0.08  &   $+$  0.08  &        Palomar 1.2-m \\
    932.2942 &     19.42  &   $-$  0.10  &   $+$  0.09  &        Palomar 1.2-m \\
\enddata
\label{tab:2002ic_lightcurve}
\end{deluxetable}

\begin{deluxetable}{lll}
\tablewidth{0pc}
\tablecaption{The V-band magnitudes for SN~2002ic as synthesized from the VLT spectrophotometry of \citet{wang04} and shown in Fig.~\ref{fig:2002ic_template}.}
\tablehead{
\colhead{JD - 2452000} & \colhead{V Mag} & \colhead{V Mag}       \\
\colhead{}             & \colhead{}      & \colhead{Uncertainty} 
}
\startdata
     829 &     19.05  &   $\pm$  0.05  \\
     850 &     19.22  &   $\pm$  0.05  \\
     852 &     19.15  &   $\pm$  0.10  \\
     912 &     19.30  &   $\pm$  0.05  \\
\enddata
\label{tab:2002ic_VLT}
\end{deluxetable}

\clearpage

\clearpage

\addcontentsline{toc}{section}{References}
\begin{thebibliography}{40}
\expandafter\ifx\csname natexlab\endcsname\relax\def\natexlab#1{#1}\fi

\bibitem[{Braun(2003)}]{hpwren}
Braun, H.-W. 2003, {High Performance Wireless Research and Education Network},
  {\em http://hpwren.ucsd.edu/}

\bibitem[{{Cappellaro} {et~al.}(1999){Cappellaro}, {Turatto}, \&
  {Mazzali}}]{iauc7091}
{Cappellaro}, E., {Turatto}, M., \& {Mazzali}, P. 1999, \iaucirc, 7091, 1

\bibitem[{{Chevalier} \& {Fransson}(1994)}]{chevalier94}
{Chevalier}, R.~A. \& {Fransson}, C. 1994, \apj, 420, 268

\bibitem[{{Chevalier} \& {Fransson}(2001)}]{chevalier01}
---. 2001, ArXiv Astrophysics e-prints, 0110060

\bibitem[{{Chevalier} \& {Soker}(1989)}]{chevalier89}
{Chevalier}, R.~A. \& {Soker}, N. 1989, \apj, 341, 867

\bibitem[{{Chugai} {et~al.}(2002){Chugai}, {Blinnikov}, {Fassia}, {Lundqvist},
  {Meikle}, \& {Sorokina}}]{chugai02}
{Chugai}, N.~N., {Blinnikov}, S.~I., {Fassia}, A., {Lundqvist}, P., {Meikle},
  W.~P.~S., \& {Sorokina}, E.~I. 2002, \mnras, 330, 473

\bibitem[{{Chugai} \& {Danziger}(1994)}]{chugai94}
{Chugai}, N.~N. \& {Danziger}, I.~J. 1994, \mnras, 268, 173

\bibitem[{{Chugai} \& {Yungelson}(2004)}]{chugai04}
{Chugai}, N.~N. \& {Yungelson}, L.~R. 2004, Astron. Letters, 30, 65

\bibitem[{{Cumming} {et~al.}(1996){Cumming}, {Lundqvist}, {Smith}, {Pettini},
  \& {King}}]{cumming96}
{Cumming}, R.~J., {Lundqvist}, P., {Smith}, L.~J., {Pettini}, M., \& {King},
  D.~L. 1996, \mnras, 283, 1355

\bibitem[{{Deng} {et~al.}(2004){Deng}, {Kawabata}, {Ohyama}, {Nomoto},
  {Mazzali}, {Wang}, {Jeffery}, {Iye}, {Tomita}, \& {Yoshii}}]{deng04}
{Deng}, J., {Kawabata}, K.~S., {Ohyama}, Y., {Nomoto}, K., {Mazzali}, P.~A.,
  {Wang}, L., {Jeffery}, D.~J., {Iye}, M., {Tomita}, H., \& {Yoshii}, Y. 2004,
  \apjl, 605, L37

\bibitem[{{Dwarkadas} \& {Chevalier}(1998)}]{dwarkadas98}
{Dwarkadas}, V.~V. \& {Chevalier}, R.~A. 1998, \apj, 497, 807

\bibitem[{{Filippenko}(1997)}]{filippenko97}
{Filippenko}, A.~V. 1997, \araa, 35, 309

\bibitem[{{Freedman} {et~al.}(2001){Freedman}, {Madore}, {Gibson}, {Ferrarese},
  {Kelson}, {Sakai}, {Mould}, {Kennicutt}, {Ford}, {Graham}, {Huchra},
  {Hughes}, {Illingworth}, {Macri}, \& {Stetson}}]{freedman01}
{Freedman}, W.~L., {Madore}, B.~F., {Gibson}, B.~K., {Ferrarese}, L., {Kelson},
  D.~D., {Sakai}, S., {Mould}, J.~R., {Kennicutt}, R.~C., {Ford}, H.~C.,
  {Graham}, J.~A., {Huchra}, J.~P., {Hughes}, S.~M.~G., {Illingworth}, G.~D.,
  {Macri}, L.~M., \& {Stetson}, P.~B. 2001, \apj, 553, 47

\bibitem[{{Gerardy} {et~al.}(2004){Gerardy}, {H{\"o}flich}, {Quimby}, {Wang},
  {Wheeler}, {Fesen}, {Marion}, {Nomoto}, \& {Schaefer}}]{gerardy04}
{Gerardy}, C.~L., {H{\"o}flich}, P., {Quimby}, R., {Wang}, L., {Wheeler},
  J.~C., {Fesen}, R.~A., {Marion}, G.~H., {Nomoto}, K., \& {Schaefer}, B.~E.
  2004, \apj, {\em in press}

\bibitem[{{Germany} {et~al.}(2000){Germany}, {Reiss}, {Sadler}, {Schmidt}, \&
  {Stubbs}}]{germany00}
{Germany}, L.~M., {Reiss}, D.~J., {Sadler}, E.~M., {Schmidt}, B.~P., \&
  {Stubbs}, C.~W. 2000, \apj, 533, 320

\bibitem[{{Goldhaber} {et~al.}(2001){Goldhaber}, {Groom}, {Kim}, {Aldering},
  {Astier}, {Conley}, {Deustua}, {Ellis}, {Fabbro}, {Fruchter}, {Goobar},
  {Hook}, {Irwin}, {Kim}, {Knop}, {Lidman}, {McMahon}, {Nugent}, {Pain},
  {Panagia}, {Pennypacker}, {Perlmutter}, {Ruiz-Lapuente}, {Schaefer},
  {Walton}, \& {York}}]{goldhaber01}
{Goldhaber}, G., {Groom}, D.~E., {Kim}, A., {Aldering}, G., {Astier}, P.,
  {Conley}, A., {Deustua}, S.~E., {Ellis}, R., {Fabbro}, S., {Fruchter}, A.~S.,
  {Goobar}, A., {Hook}, I., {Irwin}, M., {Kim}, M., {Knop}, R.~A., {Lidman},
  C., {McMahon}, R., {Nugent}, P.~E., {Pain}, R., {Panagia}, N., {Pennypacker},
  C.~R., {Perlmutter}, S., {Ruiz-Lapuente}, P., {Schaefer}, B., {Walton},
  N.~A., \& {York}, T. 2001, \apj, 558, 359

\bibitem[{{Hamuy} {et~al.}(2002){Hamuy}, {Maza}, \& {Phillips}}]{iauc8028}
{Hamuy}, M., {Maza}, J., \& {Phillips}, M. 2002, \iaucirc, 8028, 2

\bibitem[{{Hamuy} {et~al.}(2003){Hamuy}, {Phillips}, {Suntzeff}, \&
  {Maza}}]{iauc8151}
{Hamuy}, M., {Phillips}, M., {Suntzeff}, N., \& {Maza}, J. 2003, \iaucirc,
  8151, 2

\bibitem[{Hamuy {et~al.}(2003)Hamuy, Phillips, Suntzeff, Maza, {Gonz\'alez},
  Roth, Krisciunas, Morrell, Green, Persson, \& McCarthy}]{hamuy03b}
Hamuy, M., Phillips, M.~M., Suntzeff, N.~B., Maza, J., {Gonz\'alez}, L.~E.,
  Roth, M., Krisciunas, K., Morrell, N., Green, E.~M., Persson, S.~E., \&
  McCarthy, P.~J. 2003, Nature, 424, 651

\bibitem[{{Herpin} {et~al.}(2002){Herpin}, {Goicoechea}, {Pardo}, \&
  {Cernicharo}}]{herpin02}
{Herpin}, F., {Goicoechea}, J.~R., {Pardo}, J.~R., \& {Cernicharo}, J. 2002,
  \apj, 577, 961

\bibitem[{{Knop} {et~al.}(2003){Knop}, {Aldering}, {Amanullah}, {Astier},
  {Blanc}, {Burns}, {Conley}, {Deustua}, {Doi}, {Ellis}, {Fabbro}, {Folatelli},
  {Fruchter}, {Garavini}, {Garmond}, {Garton}, {Gibbons}, {Goldhaber},
  {Goobar}, {Groom}, {Hardin}, {Hook}, {Howell}, {Kim}, {Lee}, {Lidman},
  {Mendez}, {Nobili}, {Nugent}, {Pain}, {Panagia}, {Pennypacker}, {Perlmutter},
  {Quimby}, {Raux}, {Regnault}, {Ruiz-Lapuente}, {Sainton}, {Schaefer},
  {Schahmaneche}, {Smith}, {Spadafora}, {Stanishev}, {Sullivan}, {Walton},
  {Wang}, {Wood-Vasey}, \& {Yasuda}}]{knop03}
{Knop}, R.~A., {Aldering}, G., {Amanullah}, R., {Astier}, P., {Blanc}, G.,
  {Burns}, M.~S., {Conley}, A., {Deustua}, S.~E., {Doi}, M., {Ellis}, R.,
  {Fabbro}, S., {Folatelli}, G., {Fruchter}, A.~S., {Garavini}, G., {Garmond},
  S., {Garton}, K., {Gibbons}, R., {Goldhaber}, G., {Goobar}, A., {Groom},
  D.~E., {Hardin}, D., {Hook}, I., {Howell}, D.~A., {Kim}, A.~G., {Lee}, B.~C.,
  {Lidman}, C., {Mendez}, J., {Nobili}, S., {Nugent}, P.~E., {Pain}, R.,
  {Panagia}, N., {Pennypacker}, C.~R., {Perlmutter}, S., {Quimby}, R., {Raux},
  J., {Regnault}, N., {Ruiz-Lapuente}, P., {Sainton}, G., {Schaefer}, B.,
  {Schahmaneche}, K., {Smith}, E., {Spadafora}, A.~L., {Stanishev}, V.,
  {Sullivan}, M., {Walton}, N.~A., {Wang}, L., {Wood-Vasey}, W.~M., \&
  {Yasuda}, N. 2003, \apj, 598, 102

\bibitem[{{Kwok}(1993)}]{kwok93}
{Kwok}, S. 1993, \araa, 31, 63

\bibitem[{{Matzner} \& {McKee}(1999)}]{matzner99}
{Matzner}, C.~D. \& {McKee}, C.~F. 1999, \apj, 510, 379

\bibitem[{Monet {et~al.}(1996)Monet, Bird, Canzian, Harris, Reid, Rhodes, Sell,
  Ables, Dahn, Guetter, Henden, Leggett, Levison, Luginbuhl, Martini, Monet,
  Pier, Riepe, Stone, Vrba, \& Walker}]{usnoa1}
Monet, D., Bird, A., Canzian, B., Harris, H., Reid, N., Rhodes, A., Sell, S.,
  Ables, H., Dahn, C., Guetter, H., Henden, A., Leggett, S., Levison, H.,
  Luginbuhl, C., Martini, J., Monet, A., Pier, J., Riepe, B., Stone, R., Vrba,
  F., \& Walker, R. 1996, USNO-SA1.0 Catalog (U.S. Naval Observatory,
  Washington DC)

\bibitem[{{Perlmutter} {et~al.}(1997){Perlmutter}, {Gabi}, {Goldhaber},
  {Goobar}, {Groom}, {Hook}, {Kim}, {Kim}, {Lee}, {Pain}, {Pennypacker},
  {Small}, {Ellis}, {McMahon}, {Boyle}, {Bunclark}, {Carter}, {Irwin},
  {Glazebrook}, {Newberg}, {Filippenko}, {Matheson}, {Dopita}, {Couch}, \& {The
  Supernova Cosmology Project}}]{perlmutter97b}
{Perlmutter}, S., {Gabi}, S., {Goldhaber}, G., {Goobar}, A., {Groom}, D.~E.,
  {Hook}, I.~M., {Kim}, A.~G., {Kim}, M.~Y., {Lee}, J.~C., {Pain}, R.,
  {Pennypacker}, C.~R., {Small}, I.~A., {Ellis}, R.~S., {McMahon}, R.~G.,
  {Boyle}, B.~J., {Bunclark}, P.~S., {Carter}, D., {Irwin}, M.~J.,
  {Glazebrook}, K., {Newberg}, H.~J.~M., {Filippenko}, A.~V., {Matheson}, T.,
  {Dopita}, M., {Couch}, W.~J., \& {The Supernova Cosmology Project}. 1997,
  \apj, 483, 565

\bibitem[{{Pollas} {et~al.}(1988){Pollas}, {Cappellaro}, {Turatto}, \&
  {Candeo}}]{iauc4691}
{Pollas}, C., {Cappellaro}, E., {Turatto}, M., \& {Candeo}, G. 1988, \iaucirc,
  4691, 1

\bibitem[{{Pravdo} {et~al.}(1999){Pravdo}, {Rabinowitz}, {Helin}, {Lawrence},
  {Bambery}, {Clark}, {Groom}, {Levin}, {Lorre}, {Shaklan}, {Kervin},
  {Africano}, {Sydney}, \& {Soohoo}}]{pravdo99}
{Pravdo}, S.~H., {Rabinowitz}, D.~L., {Helin}, E.~F., {Lawrence}, K.~J.,
  {Bambery}, R.~J., {Clark}, C.~C., {Groom}, S.~L., {Levin}, S., {Lorre}, J.,
  {Shaklan}, S.~B., {Kervin}, P., {Africano}, J.~A., {Sydney}, P., \& {Soohoo},
  V. 1999, \aj, 117, 1616

\bibitem[{{Rigon} {et~al.}(2003){Rigon}, {Turatto}, {Benetti}, {Pastorello},
  {Cappellaro}, {Aretxaga}, {Vega}, {Chavushyan}, {Patat}, {Danziger}, \&
  {Salvo}}]{rigon03}
{Rigon}, L., {Turatto}, M., {Benetti}, S., {Pastorello}, A., {Cappellaro}, E.,
  {Aretxaga}, I., {Vega}, O., {Chavushyan}, V., {Patat}, F., {Danziger}, I.~J.,
  \& {Salvo}, M. 2003, \mnras, 340, 191

\bibitem[{{Sabine} {et~al.}(1997){Sabine}, {Baines}, \& {Howard}}]{iauc6706}
{Sabine}, S., {Baines}, D., \& {Howard}, J. 1997, \iaucirc, 6706, 1

\bibitem[{{Siloti} {et~al.}(2000){Siloti}, {Schlegel}, {Challis}, {Jha},
  {Kirshner}, \& {Garnavich}}]{siloti00}
{Siloti}, S.~Z., {Schlegel}, E.~M., {Challis}, P., {Jha}, S., {Kirshner},
  R.~P., \& {Garnavich}, P. 2000, \baas, 32, 1538

\bibitem[{{Soker} \& {Rappaport}(2000)}]{soker00}
{Soker}, N. \& {Rappaport}, S. 2000, \apj, 538, 241

\bibitem[{{Stathakis} \& {Sadler}(1991)}]{stathakis91}
{Stathakis}, R.~A. \& {Sadler}, E.~M. 1991, \mnras, 250, 786

\bibitem[{{Turatto} {et~al.}(2000){Turatto}, {Suzuki}, {Mazzali}, {Benetti},
  {Cappellaro}, {Danziger}, {Nomoto}, {Nakamura}, {Young}, \&
  {Patat}}]{turatto00}
{Turatto}, M., {Suzuki}, T., {Mazzali}, P.~A., {Benetti}, S., {Cappellaro}, E.,
  {Danziger}, I.~J., {Nomoto}, K., {Nakamura}, T., {Young}, T.~R., \& {Patat},
  F. 2000, \apjl, 534, L57

\bibitem[{{U.S. Department of Energy}(2004)}]{esnet}
{U.S. Department of Energy}. 2004, {Energy Sciences Network}, {\em
  http://www.es.net/}

\bibitem[{{Wang} {et~al.}(2003){Wang}, {Baade}, {H{\" o}flich}, {Khokhlov},
  {Wheeler}, {Kasen}, {Nugent}, {Perlmutter}, {Fransson}, \&
  {Lundqvist}}]{wang03a}
{Wang}, L., {Baade}, D., {H{\" o}flich}, P., {Khokhlov}, A., {Wheeler}, J.~C.,
  {Kasen}, D., {Nugent}, P.~E., {Perlmutter}, S., {Fransson}, C., \&
  {Lundqvist}, P. 2003, \apj, 591, 1110

\bibitem[{{Wang} {et~al.}(2004){Wang}, {Baade}, {H{\" o}flich}, {Wheeler},
  {Kawabata}, \& {Nomoto}}]{wang04}
{Wang}, L., {Baade}, D., {H{\" o}flich}, P., {Wheeler}, J.~C., {Kawabata}, K.,
  \& {Nomoto}, K. 2004, \apjl, 604, L53

\bibitem[{{Wheeler} \& {Harkness}(1990)}]{wheeler90}
{Wheeler}, J.~C. \& {Harkness}, R.~P. 1990, Reports of Progress in Physics, 53,
  1467

\bibitem[{{Wood-Vasey} {et~al.}(2004){Wood-Vasey}, {Aldering}, {Lee}, {Loken},
  {Nugent}, {Perlmutter}, {Siegrist}, {Wang}, {Antilogus}, {Astier}, {Hardin},
  {Pain}, {Copin}, {Smadja}, {Gangler}, {Castera}, {Adam}, {Bacon},
  {Lemonnier}, {P{\' e}contal}, {P{\' e}contal}, \& {Kessler}}]{wood-vasey04}
{Wood-Vasey}, W.~M., {Aldering}, G., {Lee}, B.~C., {Loken}, S., {Nugent}, P.,
  {Perlmutter}, S., {Siegrist}, J., {Wang}, L., {Antilogus}, P., {Astier}, P.,
  {Hardin}, D., {Pain}, R., {Copin}, Y., {Smadja}, G., {Gangler}, E.,
  {Castera}, A., {Adam}, G., {Bacon}, R., {Lemonnier}, J.-P., {P{\' e}contal},
  A., {P{\' e}contal}, E., \& {Kessler}, R. 2004, New Astronomy Review, 48, 637

\bibitem[{{Wood-Vasey} {et~al.}(2002){Wood-Vasey}, {Aldering}, \&
  {Nugent}}]{iauc8019b}
{Wood-Vasey}, W.~M., {Aldering}, G., \& {Nugent}. 2002, \iaucirc, 8019, 2

\bibitem[{{Young} {et~al.}(1992){Young}, {Serabyn}, {Phillips}, {Knapp},
  {Guesten}, \& {Schulz}}]{young92}
{Young}, K., {Serabyn}, G., {Phillips}, T.~G., {Knapp}, G.~R., {Guesten}, R.,
  \& {Schulz}, A. 1992, \apj, 385, 265

\end{thebibliography}
\end{document}